*In memory of Alexander Smirnitskiy*

# Long-term operation of a multi-channel cosmic muon system based on scintillation counters with MRS APD light readout

A. Akindinov[a], V. Golovin[b], E. Grigoriev[a,c], Yu. Grishuk[a], S. Kuleshov[a,d], D. Mal'kevich[a], A. Martemiyanov[a], A. Nedosekin[a], M. Ryabinin[a], K. Voloshin[a],

[a]*Institute for Theoretical and Experimental Physics (ITEP), B. Cheremushkinskaya 25, Moscow, 117218, Russia*

[b]*Center of Perspective Technologies and Apparatus (CPTA), Preobrazhenskaya pl. 6/8, Moscow 107076, Russia*

[c]*University of Geneva, CMU, rue Michel-Servet 1, Genève 4, 1211, Switzerland*

[d]*Departamento de Física y Centro de Estudios Subatómicos, Universidad Técnica Federico Santa María, Casilla 110-V, Valparaíso, Chile*

**Abstract**

A Cosmic Ray Test Facility (CRTF) is the first large-scale implementation of a scintillation triggering system based on a new scintillation technique known as START. In START, the scintillation light is collected and transported by WLS optical fibers, while light detection is performed by pairs of avalanche photodiodes with the Metal–Resistor–Semiconductor structure operated in the Geiger mode (MRS APD). START delivers 100% efficiency of cosmic muon detection, while its intrinsic noise level is less than $10^{-2}$ Hz. CRTF, consisting of 160 START channels, has been continuously operated by the ALICE TOF collaboration for more than 25 000 hours, and has demonstrated a high level of stability. Fewer than 10% of MRS APDs had to be replaced during this period.

## 1. Introduction

During the last 15 years, a significant breakthrough has been achieved in the development of micro-cell avalanche photodiodes operated in the Geiger mode for the purpose of visible light detection [1]. Avalanche photodiodes with the Metal–Resistor–Semiconductor structure (MRS APD), developed and produced by CPTA[1], have proved to be a superior solution for Geiger-mode photodetectors with $n^{+}$–$p$–$\pi$–$p^{++}$ structure, optimized for the detection of light in the green-red spectrum region [2]. A special meso-structure with optical decoupling of micro-cells [3] ensures that the single-photon detection efficiency of MRS APD is more than 25%, while its noise measured at the level of 4 photo-electrons is about 1 kHz. These characteristics, together with the low bias voltage (20–50 V), low price, tiny dimensions (sensitive area of the order of 1 mm$^2$), and insensitivity to the magnetic field, make MRS APD a competitive device as compared to multi-anode PMT [1].

Initially developed for hadron calorimetry, the method of light collection by means of WLS fibers, positioned in ring-shaped grooves inside organic scintillating plastic plates, has been imple-mented in a scintillation counter with MRS APD light readout (START). The basic idea of START operation consists of simultaneous detection of light by two MRS APDs coupled to the opposite sides of a fiber piece [4].

[1] For the address, see the authors affiliation list.

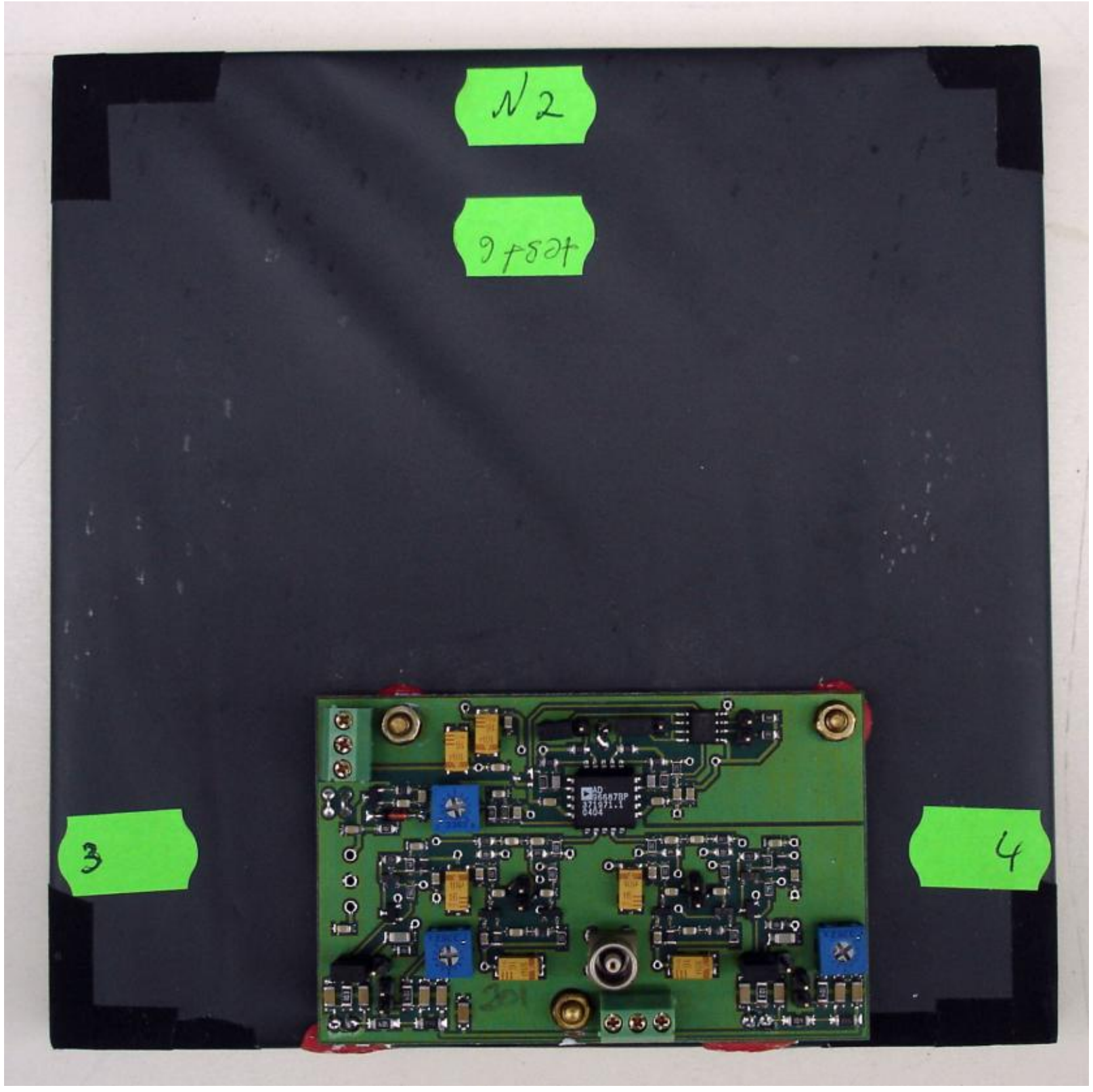

Fig. 1. General layout of START (external dimensions 15 × 15 × 1.5 $cm^3$).

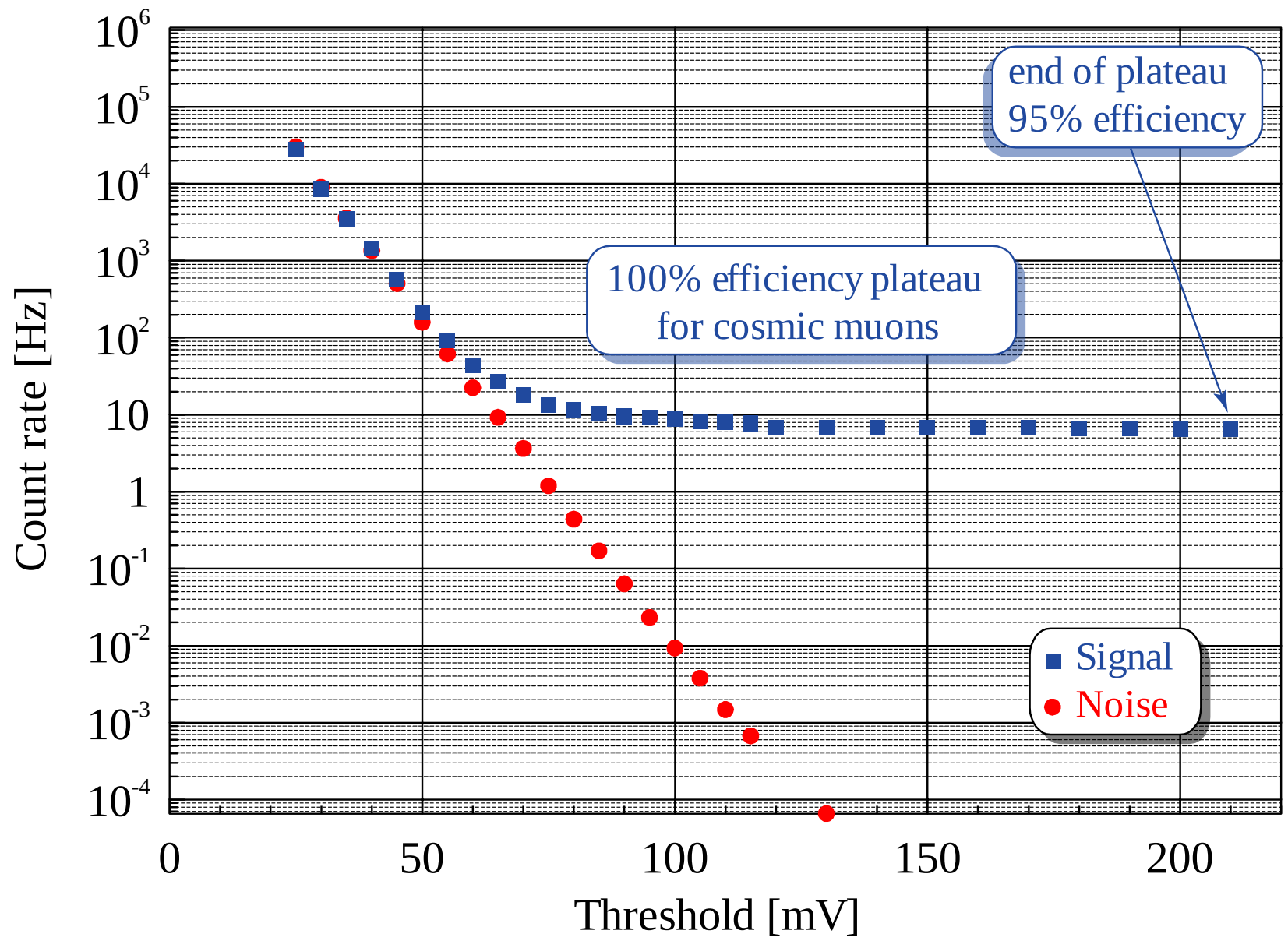

Fig. 2. Signal and noise rates of START in response to cosmic radiation.

[2] Produced by Polimersintez, Vladimir, Russia.

The rectangular shape of START allows for coverage of large areas with multi-channel arrays of such devices. No degradation of START characteristics has been observed in the first 32-channel prototype [5]. Subsequent improvement of the MRS APD performance has resulted in less stringent requirements for the quality of scintillating plastic, making the usage of relatively cheap Polisterol-165[2] possible in large-scale START implementations. This version of START was adopted as a basic element of a cosmic ray test facility (CRTF), intended for mass tests of the ALICE TOF modules [6].

## 2. START characteristics

With a compact electronic card that produces coincidence signal from the two MRS APDs and is mounted directly on its body (see photo in Fig. 1), START represents an autonomous scintillation counter.

To measure the noise characteristics and efficiency of START in response to cosmic particles, a tested START sample, sized 15 × 15 × 1.5 $cm^3$, was placed in the center between two scintillation counters with plastic dimensions of 14 × 14 × 1 $cm^3$, positioned 1.7 m apart. Coincidence of signals from the scintillation counters was used to trigger the passage of ionizing particles through START. Signals coming from the two MRS APDs were fed to discriminators with variable thresholds, which produced 50 ns-long NIM pulses. These pulses were then sent to two coincidence circuits with time gates of 100 ns, one of the pulses being delayed for 250 ns before arrival at the second circuit. Thus, the first circuit triggered real coincidences (true START signals), while the second one allowed monitoring of accidental coincidences (false START signals, or noise). Individual noise rates of the MRS APDs, $F_1$ and $F_2$, were measured as well. Changes in the threshold values were applied to both discriminators simultaneously. The rate of accidental coincidences was found to be equal to $F_1 F_2 \times 100$ ns within 5% of accuracy at all thresholds, which proved the absence of cross-talks between two photo-diode channels. Shown in Fig. 2 are the rates of true and false START events for different discriminator thresholds. When the threshold exceeds 100 mV (which is above the fourth photoelectron peak) the noise rate becomes less than $10^{-2}$ Hz, while the signal rate flattens to a quasi-plateau of 8–10 Hz, this value being consistent with the intensity of cosmic radiation.

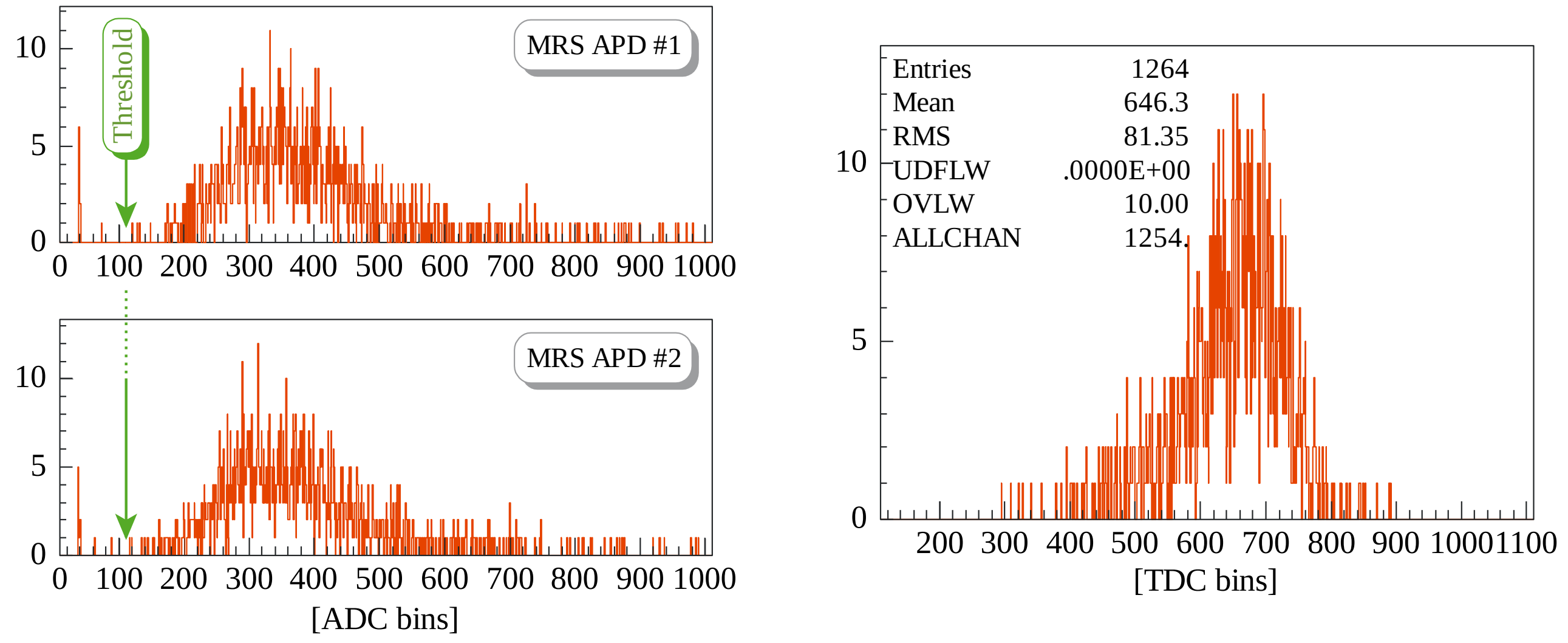

(a) (b)

Fig. 3. (a) Amplitude and (b) timing distribution of START signals, measured at the discriminator threshold of 115 mV.

Amplitude spectra of the two MRS APDs and a timing spectrum of START measured relative to the trigger produced by the scintillation counters are shown in Fig. 3. After cutting off the pedestal events (not related to START and probably induced by high-voltage equipment positioned close to the setup), the efficiency of START was evaluated by counting the number of events accumulated in the overflow TDC channel. With the discriminating threshold set at 115 mV (which is the case for the data in Fig. 3) the efficiency was found to exceed 99%. It stays this high within a wide range of discriminating thresholds, corresponding to the quasi-plateau in Fig. 2. Precise adjustment of the thresholds for different START samples is hence not necessary, and large arrays of STARTs may be put into action without complicated calibration.

## 3. Triggering system for CRTF

The TOF system of the ALICE experiment is based on more than 159 000 strips of Multi-Gap Resistive Plate Chambers (MRPC), covering the area of about 150 m$^2$ [6]. To simplify the construction and tests of the system, the MRPC strips are assembled in modules of three types: central (1.2 × 1.6 m$^2$), intermediate (1.5 × 1.6 m$^2$) and external (1.7 × 1.6 m$^2$). These modules are in turn assembled in 18 supermodules, 5 modules in each (external–intermediate–central–intermediate–external). Beam tests of the ALICE TOF modules turned out to be very time- and cost-consuming and were omitted. As an alternative, tests with rays of cosmic muons were suggested, in which up to 6 TOF modules were to be positioned over each other between two trigger layers to form a telescope. The trigger signal was assumed to be produced by coincidence of signals from the upper and lower trigger layers. The size, position and granularity of the trigger layers had to be chosen so as to provide sufficient detection and tracking efficiency, the latter being determined by the size of a single MRPC cell ($3.5 \times 2.5$ cm$^2$) and the geometry of the test facility. The triggering detectors were required to possess 100% detection efficiency towards MIP and minimum noise level. For many reasons, some of which may be found in [5], a $15 \times 15$ cm$^2$ version of START was chosen as a muon triggering counter for CRTF.

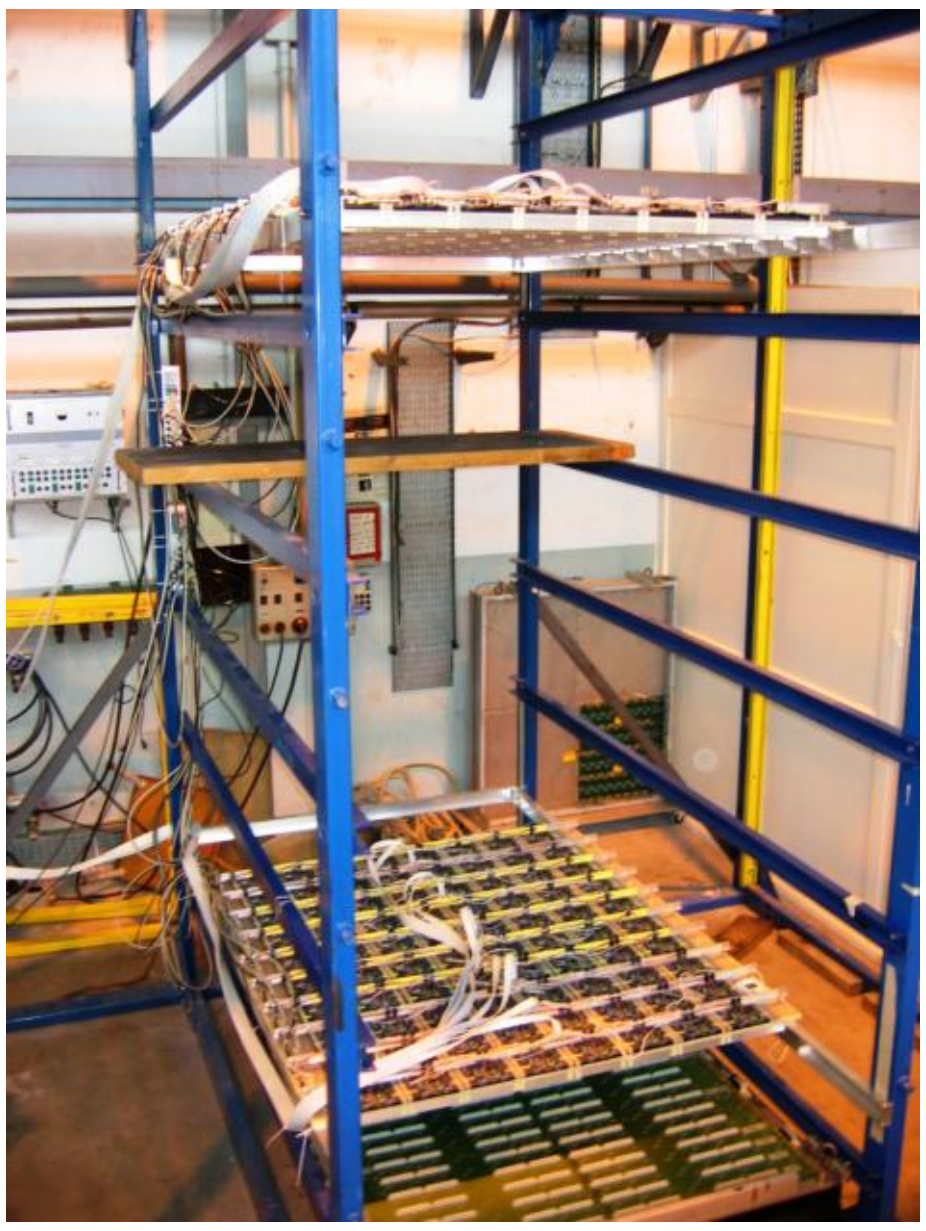

(a)

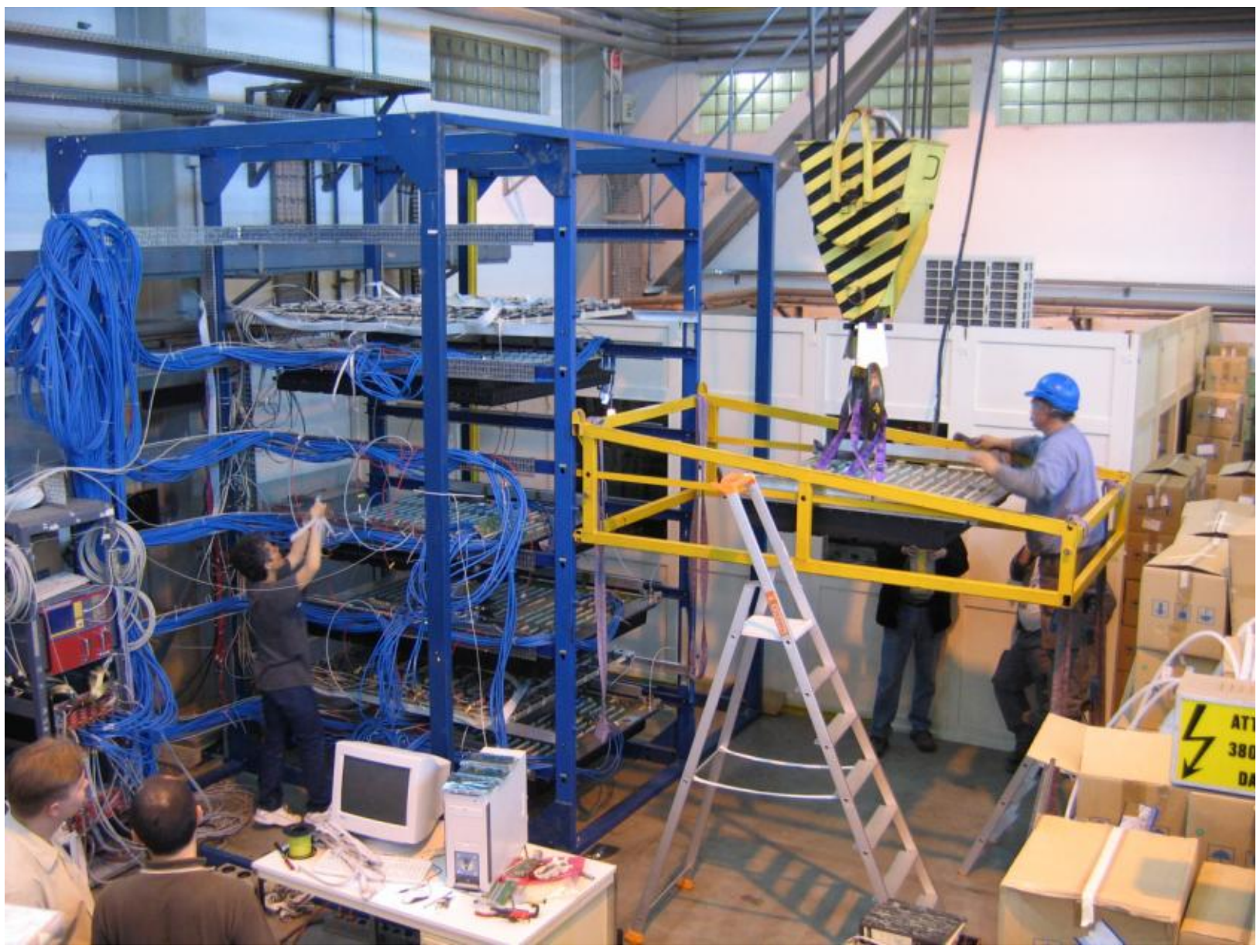

(b)

Fig. 4. CRTF design and operation: (a) steel support frame with the two START trigger layers, (b) installation of an MRPC module into the facility.

As the photo in Fig. 4a illustrates, each trigger layer ($1.2 \times 1.5$ m$^2$) consisted of 80 STARTs, combined in 10 strips, 8 detectors per strip. The distance between the two trigger layers was about 2.3 m. The total number of STARTs was therefore 160, and the number of MRS APDs equaled to 320. Bias voltages for individual MRS APDs could be adjusted by resistive dividers positioned on electronic boards, which made it possible to use a single power supply unit for all photodiodes. The rest of the electronics were powered with two identical power supply units.

The whole setup was assembled on a steel support frame. Insertion of TOF modules was facilitated by using rollers and horizontal rails positioned on both sides of the frame, as well as a special movable shelf shown on photo in Fig. 4b.

## 4. Trigger configuration and efficiency monitoring

Since the outputs of electronic cards were logical signals, the trigger formation and system monitoring were implemented by programmable logical blocks, controlled with the Labview[3] package by means of a Wienner[4] CC-USB controller. Fig. 5a shows the front panel of the trigger configuration program. Any counter could be included or excluded from the coincidence. This was helpful if, for instance, any part of a TOF module was not functioning well and had to be repaired: The trigger schematics made the study of a particular part of the module area possible in repeated tests.

[3] National Instruments Corporation, Austin, Texas, www.ni.com .

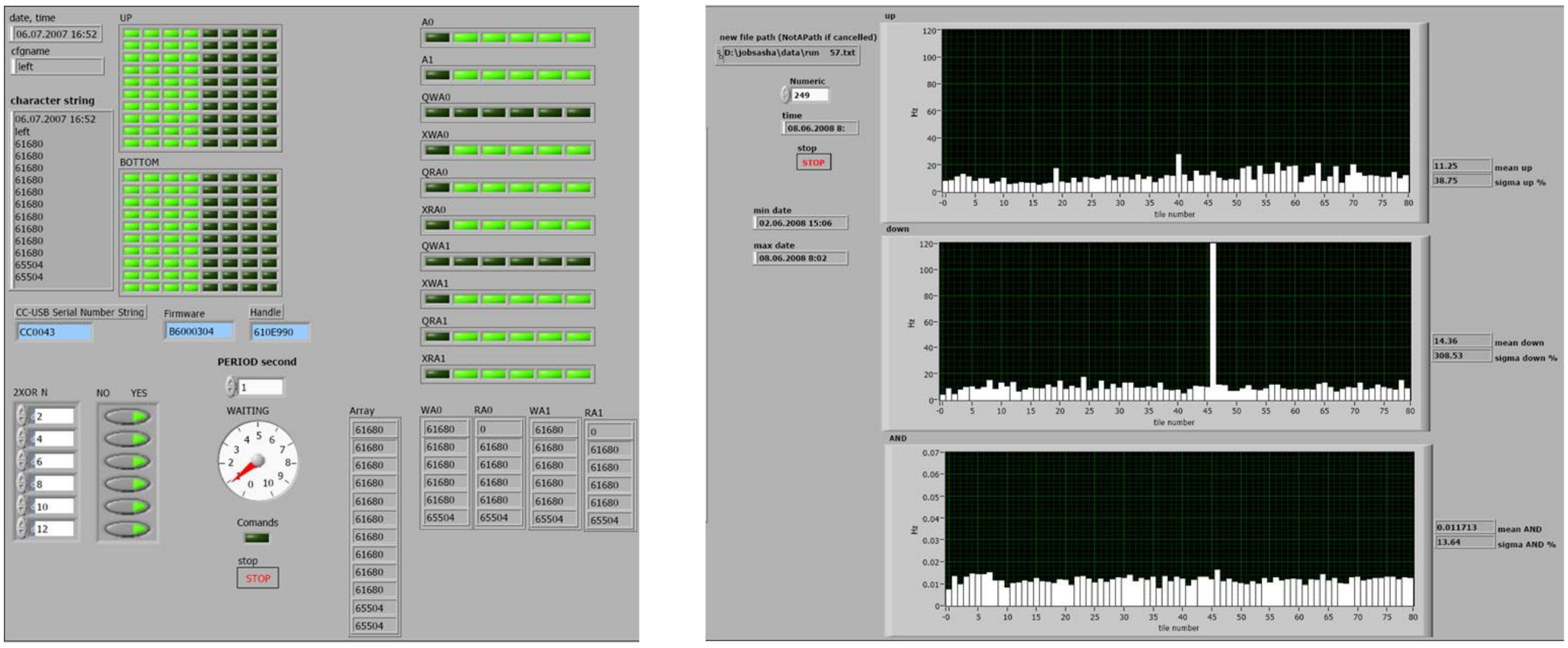

(a) (b)

Fig. 5. Front panels of the trigger controlling programs: (a) trigger configuration, (b) count rates in the trigger layers and coincidence rate.

In case a coincidence signal from the trigger layers was generated, a cosmic particle track was lined through two fired STARTs, which allowed measurement of efficiencies of the MRPC cells hit by the track. Due to low intrinsic noise of STARTs, the input of noise events into these estimations was negligible. Inefficient MRPC cells were therefore selected by the decrease of coincidence rate in the corresponding region of module surface. Presented in Fig. 5b is the front panel of the monitoring gram showing distributions of count rates in the two triggering layers and the coincidence rate.

## 5. Changes of the CRTF characteristics in the process of its exploitation

The coincidence count rates were measured at the beginning of CRTF operation in the absence of the TOF modules and in 6 months after its launch with the TOF modules present. The results are presented in Fig. 6. A visible distinction of mean values may be explained by rescattering of cosmics in the material of the TOF modules (the solid angle for each pair of STARTs is less than 2 degrees), front-end electronic cards and cable lines. Different types of TOF modules varied in length which explains the dependence of the count rate on the strip number (all modules were aligned to the left, i.e. to the channels 71–80). One can also see START pairs with significantly lower count rates which is a sign of the loss of efficiency by at least one of the two counters in a pair.

The consumption currents of MRS APDs could only be measured during short breaks in the test runs, when both trigger planes became accessible. Fig. 7 shows the relation of currents measured in 1 and 6 months after the launch of CRTF to their initial values for each MRS APD in one trigger layer. The outliers correspond to the STARTs, for which the decrease of efficiency was previously seen (bins 20, 57, 126 in Fig. 7 correspond to bins 11, 28, 63 in Fig. 6). This proves that the loss of efficiency is due to the fall of supply voltage on the protection serial resistor of 100 kΩ. After this effect was discovered, the malfunctioning STARTs were replaced during the following test interruptions. Four cases were accompanied with a complete loss of efficiency by corresponding counters; measurement of the consumption current for photodiodes in these counters proved them to be zero.

[4] WIENER, Plein & Baus GmbH, www.wiener-d.com .

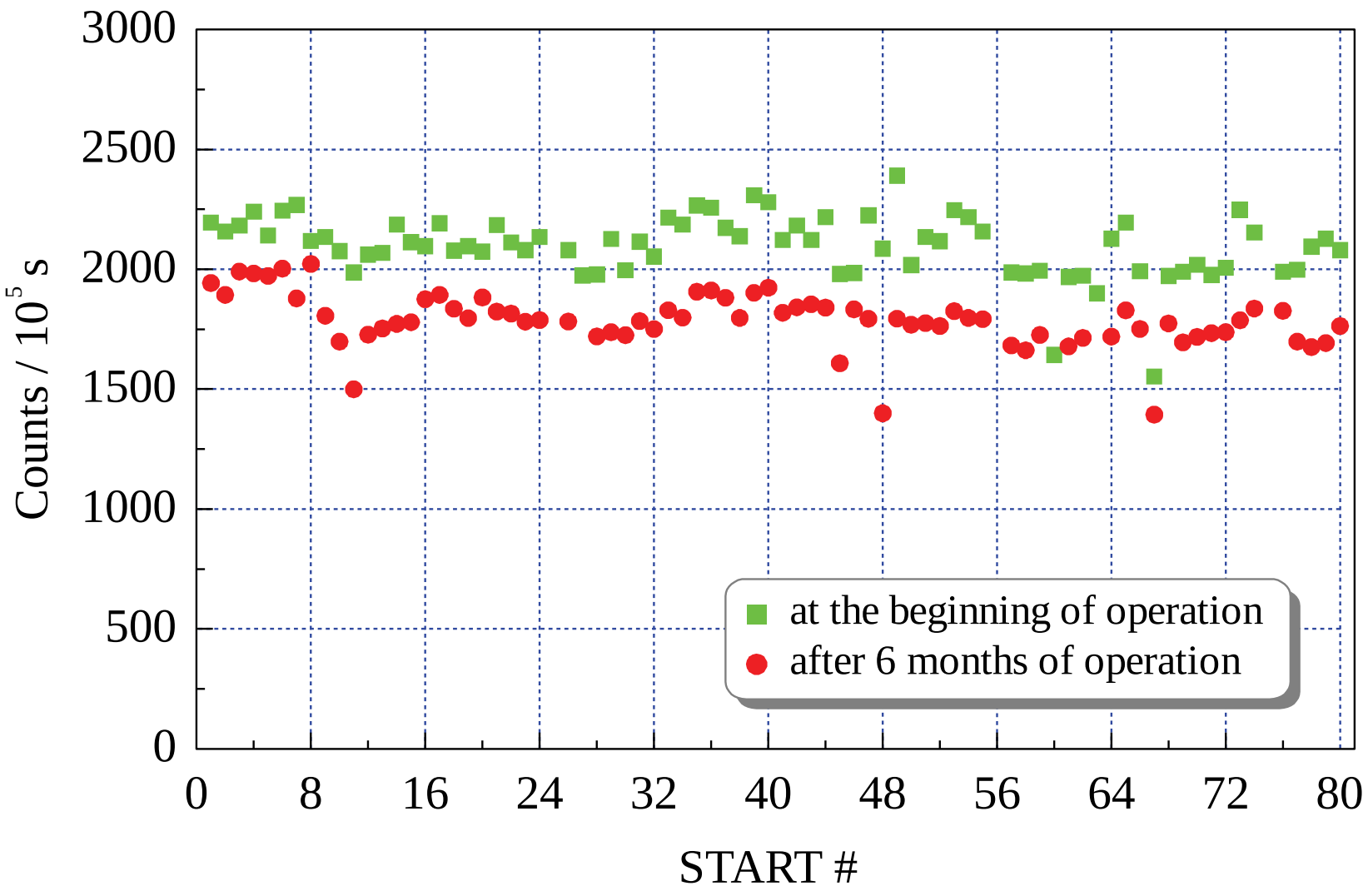

Fig. 6. Comparison of START count rates in one trigger plane before the launch of CRTF (squares) and after 6 months of tests (circles).

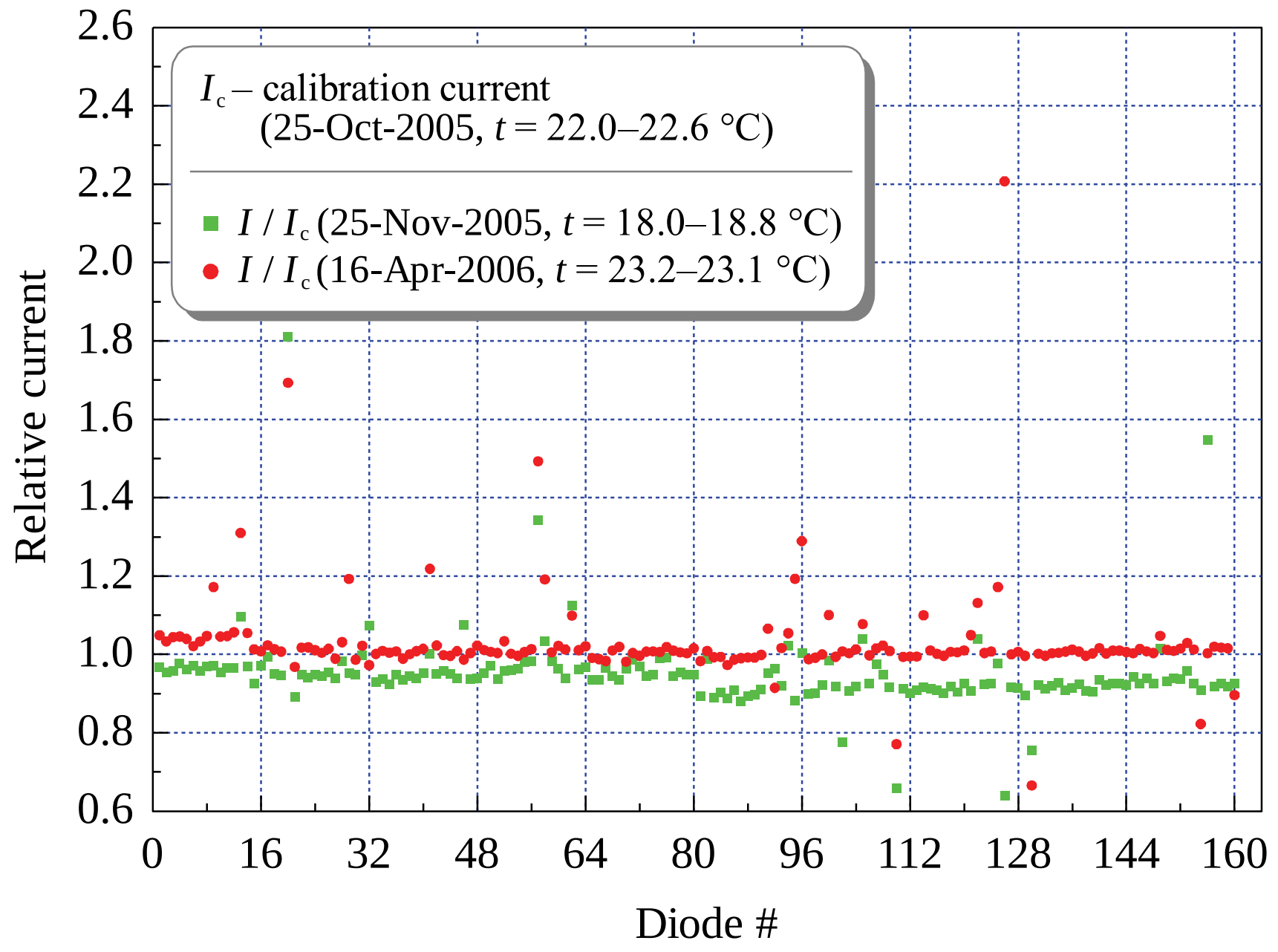

Fig. 7. Comparison of consumption currents of MRS APDs after 1 (squares) and 6 (circles) months of CRTF operation.

## 6. Study of broken MRS APDs

After malfunctioning counters were disassembled, the broken photodiodes were analyzed. For those MRS APDs with zero consumption currents, a loss of contact between the photodiode pad and the bonding wire (see Fig. 8a) was discovered. Although the number of these photodiodes is small, the producer has been informed and is currently working on the improvement of the quality of bonding.

There were a total of 27 MRS APDs that demonstrated an increased consumption current. Overtime, their consumption current continued to grow up to 20 µA and even further. The noise spectrum of these photodiodes contained a high-frequency component of more than 20 MHz, with the signal amplitudes 10 times less than those produced by single photoelectrons. It is reasonable to assume that this effect is caused by an appearance of a conductive channel between the metal and $p$-type semiconductor layers of the photodiode structure due to the thinning of the $SiO_2$ protective layer in one or several micro-cells (see Fig. 8b). CPTA is currently analyzing these results to improve the MRS APD production technology.

## 7. Conclusion

For the first time, a large-scale scintillation trigger system based on MRS APDs has been continuously exploited for a long time period of almost 3 years. As a result of this work, all 87 modules of the ALICE TOF system have been successfully tested and calibrated. The calibration results will be published in a separate paper.

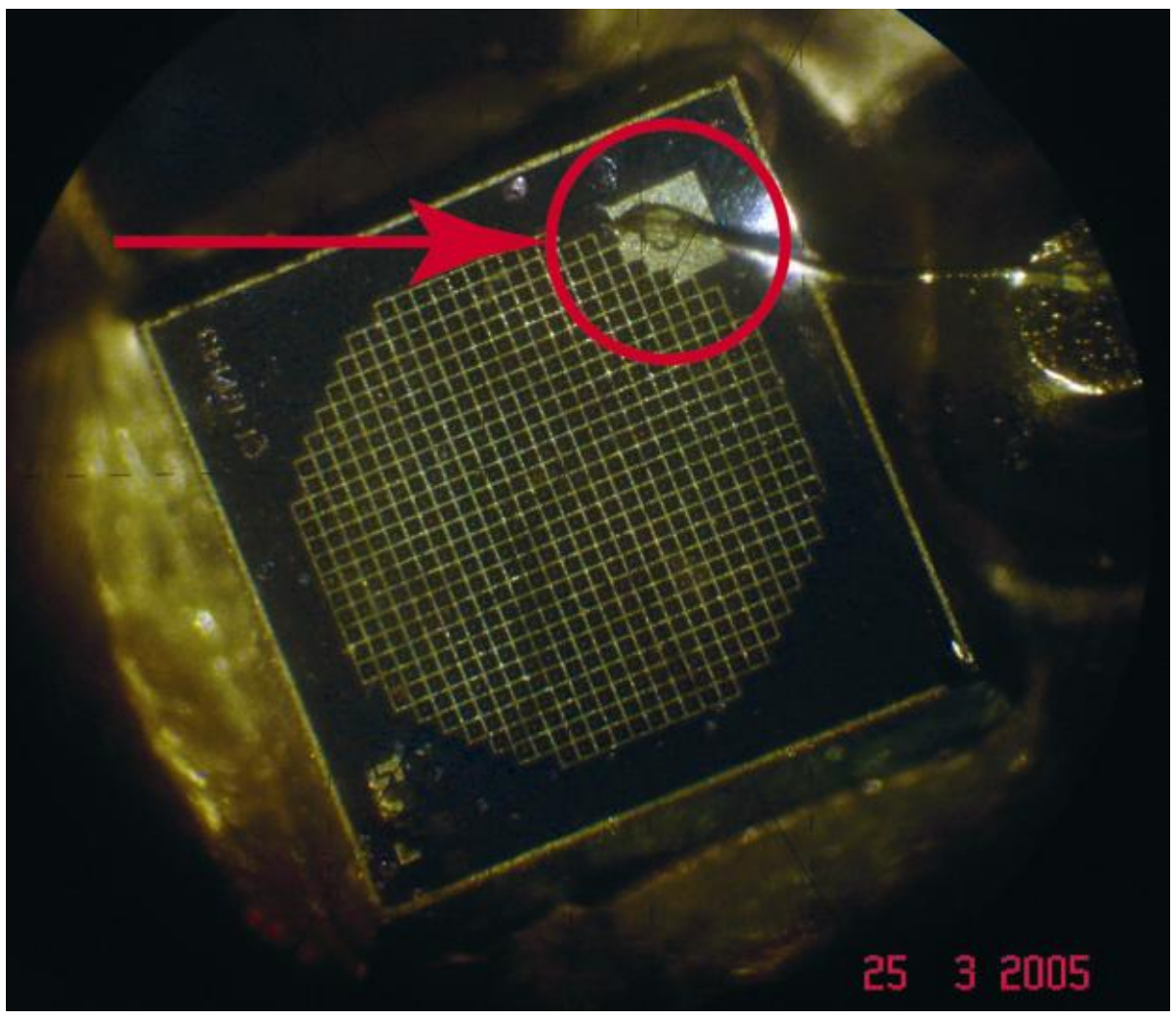

(a)

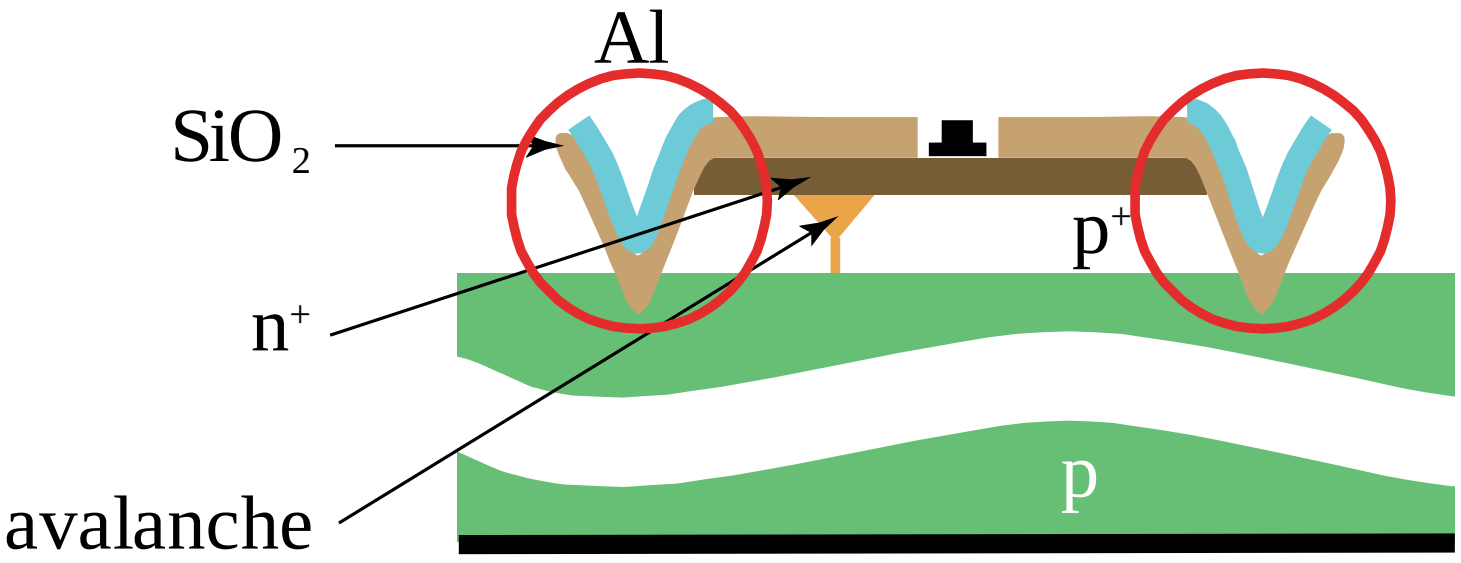

(b)

Fig. 8. Illustrations to the explanation of degradation of MRS APD characteristics (see Section 6): (a) location of the bonding wire contact on the sensitive surface of photodiode, (b) cross-section of the MRS APD structure with the possible areas of $SiO_2$ thinning marked with red circles.

ALICE TOF CRTF built of STARTs has proved to be relatively inexpensive and easy-to-operate. The two-layer design of the trigger system makes it possible to control its efficiency based on the count rate from cosmic muons without LED monitoring. Fewer than 10% of photodiodes were identified as malfunctioning during the entire testing period.

The trigger system described in this article may be exploited both in experiments on high-energy physics and in various applied sciences. For instance, the sensitivity of the system to multiple re-scattering may be useful in detection of foreign objects in a homogeneous medium (customs control) or in detection of radioactive sources of low-intensity in wagons and vehicles (security control).

## Acknowledgements

The authors are grateful to all their collaborators from INFN-Bologna and INFN-Salerno for fruitful joint work on the ALICE TOF project. Our special thanks are addressed to Despina Chatzifotiadou who actively participated in the tests at CRTF.

This work was partially supported from the RFBR grant 08–02–91011–CERN_a and from the funding program LHC-2 of the Russian Federal Agency of Atomic Energy (Rosatom).

## References

[1] Y. Musienko, S. Reucroft and J. Swain, Nucl. Instr. and Meth. A 567 (2006) 57.

[2] A. Akindinov, G. Bondarenko, V. Golovin, et al., Nucl. Instr. And Meth. A 567 (2006) 74.

[3] V. Golovin, G. Bondarenko and M. Tarasov, Patent of Russia № 2142175, http://www.sibpatent.ru/patent.asp?nPubl=2142175&mpkcls=H01L031&ptncls=H01L031/06&sort=2 (in Russian).

[4] A. Akindinov, G. Bondarenko, V. Golovin, et al., Nucl. Instr. And Meth. A 539 (2005) 172.

[5] A. Akindinov, G. Bondarenko, V. Golovin, et al., Nucl. Instr. And Meth. A 555 (2005) 65-71.

[6] ALICE Collaboration, Time-of-flight system, Addendum to ALICE TDR 8, CERN/LHCC 2002–16.